\documentclass{rnaastex}

\begin{document}

\title{A Single Radio Source in the High-redshift Dual AGN LBQS\,0302--0019}

\correspondingauthor{S\'andor Frey}
\email{frey.sandor@csfk.mta.hu}

\author[0000-0003-3079-1889]{S\'andor Frey}
\affiliation{Konkoly Observatory, MTA CSFK, Konkoly \'ut 15-17, H-1121 Budapest, Hungary}

\author{Krisztina \'E. Gab\'anyi}
\affiliation{Konkoly Observatory, MTA CSFK, Konkoly \'ut 15-17, H-1121 Budapest, Hungary}
\affiliation{MTA-ELTE Extragalactic Astrophysics Research Group, P\'azm\'any  s\'et\'any 1/A, H-1117 Budapest, Hungary}

\keywords{radio continuum: galaxies --- quasars: individual: LBQS\,0302--0019}

\section{} 

By discovering a narrow-line emitter embedded in an extended Ly$\alpha$ nebula, \citet{2018A&A...610L...7H} found that the luminous radio-quiet quasar LBQS\,0302--0019 is a dual active galactic nucleus (AGN) system. The separation of the unobscured quasar and its obscured companion is $2\farcs9$, corresponding to a projected linear distance $\sim22$~kpc, assuming a flat $\Lambda$CDM cosmological model with $H_0=70$ km\,s$^{-1}$\,Mpc$^{-1}$ and $\Omega_\mathrm{M}=0.3$. The redshift $z=3.2859$ falls in the cosmological era of intense galaxy interactions where kpc-separation dual AGNs should be ubiquitous according to the hierarchical galaxy formation paradigm.

Radio waves are free from dust obscuration and thus offer a means of detecting AGN even in an obscured environment. Moreover, owing to its fine angular resolution, the technique of radio interferometry is capable of directly resolving and imaging kpc-scale dual AGNs if both objects are radio emitters. According to the conventional classification, LBQS\,0302--0019 is radio-quiet \citep{2005ApJ...618..108B}. However, it is a known weak radio source that was extensively studied in the past \citep{1995ApJ...445...62H,2005ApJ...618..108B}. In the data archive\footnote{\href{http://archive.nrao.edu}{http://archive.nrao.edu}} of the U.S. National Radio Astronomy Observatory (NRAO) Very Large Array (VLA), there are several observations of LBQS\,0302--0019 made at 8.4~GHz with various configurations of the interferometer. Most of these data were obtained by \citet{2005ApJ...618..108B} who monitored LBQS\,0302--0019 in 1997. They found its flux density invariable with an average value 0.4~mJy. Motivated by this, we re-analysed the archival VLA measurements obtained for LBQS\,0302--0019 and combined the calibrated visibility data taken at different epochs into one single data set. This way the image noise level could be improved by a factor of $\sim$3 with respect to that achievable with the individual observations, increasing a chance of the detection of any putative weak radio emission from the obscured companion.

For the combined analysis, we considered all available archival 8.4-GHz VLA data except for those taken in the most compact D configuration of the array that is insufficient to resolve sources separated by $2\farcs9$. The VLA project codes and observation dates were the following: project AI042, 1991 Jan 29 \citep{1995ApJ...445...62H}; project AL410, 1997 Jan 18, Mar 12, May 04; project AL418, 1997 May 30, Jun 29, Aug 07, Aug 21, Sep 20 \citep{2005ApJ...618..108B}. Following \citet{2005ApJ...618..108B}, we discarded data from 1997 Apr 03 because of bad weather.

\begin{figure}[h!]
\begin{center}
\includegraphics[scale=0.5,angle=0]{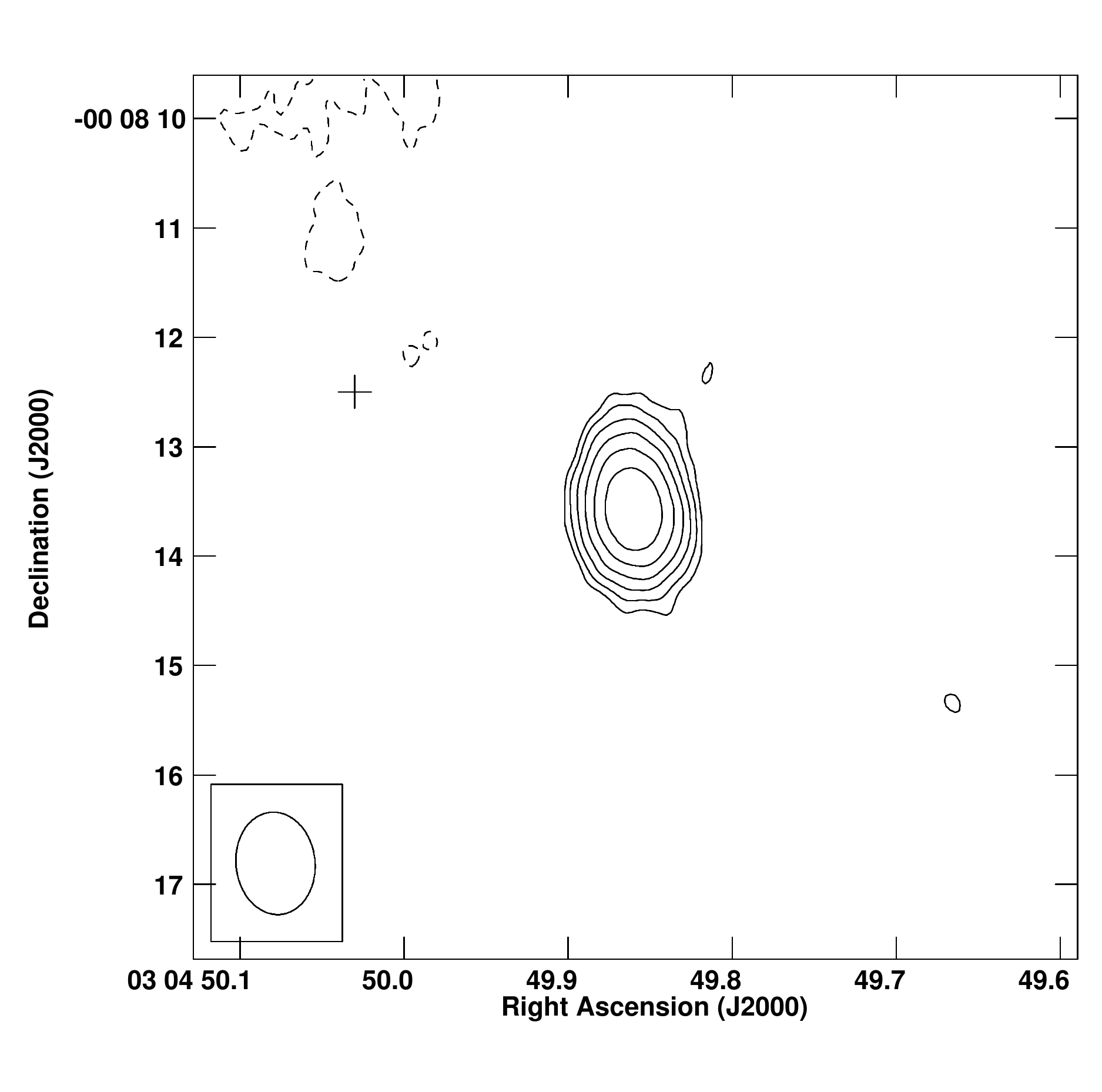}
\caption{Naturally-weighted 8.4-GHz VLA image of LBQS\,0302--0019. The peak brightness is 365~$\mu$Jy\,beam$^{-1}$, the lowest contours are $\pm$45~$\mu$Jy\,beam$^{-1}$, positive contours increase by a factor of $\sqrt2$. The restoring beam is $0\farcs94 \times 0\farcs72$ with a major axis position angle $6\fdg$2. The cross marks the location of the companion AGN \citep{2018A&A...610L...7H}.
\label{fig1}}
\end{center}
\end{figure}

The NRAO Astronomical Image Processing System \citep[AIPS,][]{2003ASSL..285..109G} was employed to calibrate the VLA data in the standard manner. Primary flux density calibrators 3C\,48, 3C\,147, and 3C\,286 were used, depending on the scheduled calibrator scans at different epochs. Each individual data set was exported to Difmap \citep{1994BAAS...26..987S} for imaging. We also fitted a circular Gaussian brightness distribution model to the visibility data and confirmed that the flux density values do not vary significantly. The measured values were between 0.37~mJy and 0.46~mJy, with typical error bars of 0.06~mJy \citep[see also][]{2005ApJ...618..108B}. The average flux density was 0.41~mJy. We present the 8.4-GHz VLA image of LBQS\,0302--0019 (Fig.~\ref{fig1}) constructed from the 9-epoch data combined in AIPS. The fitted Gaussian component flux density is 0.40$\pm$0.05~mJy.

LBQS\,0302--0019 shows a single compact radio component located at right ascension $03^{\rm h} 04^{\rm m} 49\fs861$, declination $-00{\degr} 08{\arcmin} 13\farcs57$ (J2000), with an estimated positional uncertainty $\la 0\farcs1$. This coincides with the known optical position of the (unobscured) quasar ($03^{\rm h} 04^{\rm m} 49\fs858$, $-00{\degr} 08{\arcmin} 13\farcs54$) within the errors \citep{2015A&A...583A..75S}. As seen in Fig.~\ref{fig1}, the obscured companion found by \citet{2018A&A...610L...7H} has no arcsec-scale radio emission above the brightness level $\sim45$~$\mu$Jy\,beam$^{-1}$ (3$\sigma$). However, the non-detection of this object with the VLA does not rule out the presence of a radio-quiet AGN there.

The rest-frame monochromatic 8.4-GHz radio power of LBQS\,0302--0019 is $(5.7\pm1.4) \times 10^{24}$ W\,Hz$^{-1}$ which may be partly related to AGN and star formation in the host galaxy. For $K$-correction, we assumed the radio spectral index +0.19 \citep{2005ApJ...618..108B}. This flat spectrum suggests that some radio emission may originate from synchrotron jets. To better constrain the radio emission mechanism in LBQS\,0302--0019, milli-arcsec resolution very long baseline interferometry observations are desirable.

In summary, the radio source in this system positionally coincides with the unobscured optical quasar while its obscured companion \citep{2018A&A...610L...7H} has no detectable radio emission.

\acknowledgments

This work was supported by the NKFIH-OTKA\,NN110333 grant. K.\'E.G. acknowledges the Bolyai Research Scholarship of the Hungarian Academy of Sciences. The NRAO is a facility of the National Science Foundation operated under cooperative agreement by Associated Universities, Inc.

\end{document}